\pdfoutput=1
\documentclass[a4paper,11pt,showpacs,amsmath,nofootinbib,superscriptaddress]{article}
\usepackage{jcappub.local}

\usepackage{subfigure}
\usepackage[subfigure]{graphfig}
\usepackage{color}
\usepackage[toc,page]{appendix}

\usepackage{graphicx}
\usepackage{comment}
\usepackage{array,amsmath,amsthm,feynmf}
\usepackage{bm}
\usepackage{slashed}
\usepackage{times}
\usepackage{epsfig}
\usepackage{cancel}
\usepackage{amssymb}
\usepackage{textcomp}
\usepackage{mathrsfs}

\usepackage[section]{placeins}
\makeatletter
\AtBeginDocument{%
  \expandafter\renewcommand\expandafter\subsection\expandafter{%
    \expandafter\@fb@secFB\subsection
  }%
}
\makeatother

\title{\boldmath Differentiating G-inflation from String Gas Cosmology using the Effective Field Theory Approach }

\author[a,c,f]{Minxi He,}
\author[a,d,g]{Junyu Liu,}
\author[a,c]{Shiyun Lu,}
\author[e]{Siyi Zhou,}
\author[a,b,c,*]{\\Yi-Fu Cai,}
\author[e,*]{Yi Wang,}
\author[b]{Robert Brandenberger,}

\affiliation[a]{CAS Key Laboratory for Research in Galaxies and Cosmology, Department of Astronomy, University of Science and Technology of China, Chinese Academy of Sciences, Hefei, Anhui 230026, China}
\affiliation[b]{Department of Physics, McGill University, Montr\'eal, Quebec H3A 2T8, Canada}
\affiliation[c]{School of Physical Sciences, University of Science and Technology of China, Hefei, Anhui 230026, China}
\affiliation[d]{School of the Gifted Young, University of Science and Technology of China, Hefei, Anhui 230026, China}
\affiliation[e]{Department of Physics, The Hong Kong University of Science and Technology, Clear Water Bay, Kowloon, Hong Kong, China}
\affiliation[f]{Department of Physics, Graduate School of Science, The University of Tokyo, Tokyo 113-0033, Japan}
\affiliation[g]{Department of Physics, California Institute of Technology, Pasadena, California 91125, USA}

\emailAdd{hmxz0@mail.ustc.edu.cn}
\emailAdd{jliu2@caltech.edu}
\emailAdd{shiyun@mail.ustc.edu.cn}
\emailAdd{zhousiyi1@gmail.com}

\emailAdd{yifucai@ustc.edu.cn}
\emailAdd{phyw@ust.hk}
\emailAdd{rhb@physics.mcgill.ca}

\abstract{A characteristic signature of String Gas Cosmology is primordial power spectra for scalar and tensor modes which are almost scale-invariant but with a red tilt for scalar modes but a blue tilt for tensor modes. This feature, however, can also be realized in the so-called G-inflation model, in which Horndeski operators are introduced which leads to a blue tensor tilt by softly breaking the Null Energy Condition. In this article we search for potential observational differences between these two cosmologies by performing detailed perturbation analyses based on the Effective Field Theory approach. Our results show that, although both two models produce blue tilted tensor perturbations, they behave differently in three aspects. Firstly, String Gas Cosmology predicts a specific consistency relation between the index of the scalar modes $n_s$ and that of tensor ones $n_t$, which is hard to be reproduced by G-inflation. Secondly, String Gas Cosmology typically predicts non-Gaussianities which are highly suppressed on observable scales,  while G-inflation gives rise to observationally large non-Gaussianities because the kinetic terms in the action become important during inflation. However, after finely tuning the model parameters of G-inflation it is possible to obtain a blue tensor spectrum and negligible non-Gaussianities with a degeneracy between the two models. This degeneracy can be broken by a third observable, namely the scale dependence of the nonlinearity parameter, which vanishes for G-inflation but has a blue tilt in the case of String Gas Cosmology. Therefore, we conclude that String Gas Cosmology is in principle observationally distinguishable from the single field inflationary cosmology, even allowing for modifications such as G-inflation. }

\begin{document}
\maketitle
\flushbottom

\section{Introduction} \label{sec:intro}

The cosmic microwave background (CMB) temperature anisotropies have been measured to high precision and strongly support a cosmological paradigm in which the large scale structure (LSS) of our universe evolved from primordial almost Gaussian and almost adiabatic curvature perturbations with a nearly scale-invariant power spectrum \cite{Ade:2015xua} . Experiments are now providing us with CMB temperature maps which are accurate enough such that one can extract not only their specific patterns at linear order but also constrain the amplitudes and profiles of non-Gaussian fluctuations \cite{Ade:2015ava} . To study the generation mechanism of these primordial perturbations is one of the most important topics of modern cosmology, through which one can probe very early universe scenarios.

Nowadays one can study CMB maps beyond the level of pure temperature anisotropies. Since the CMB photons are polarized, the study of CMB polarization maps could yield more information about the very early universe. In particular, the CMB polarization can be decomposed into the so-called E- and B-mode. Curvature perturbations of scalar type in the early universe lead to a pure E-mode signal. B-mode polarization can originate from gravitational waves generated in the primordial epoch, or from vector and tensor perturbations at late times, such as in topological defect models \cite{RHBCS, Durrer}. Different early universe cosmologies make rather different predictions for the spectrum of primordial gravitational waves. Hence, assuming for a moment the absence of late time vector and tensor fluctuations, the search for the B-mode signals in the CMB becomes a possible way of studying the nature of the early universe when quantum gravity effects are important. The E-mode polarization patterns have already been seen in the CMB surveys, and the results provide more evidence to support the standard paradigm of the hot big bang $\Lambda$CDM cosmology \cite{Ade:2015xua} . The B-mode signals, however, have not yet been observed in any cosmological/astronomical experiment.

According to the latest joint analysis \cite{joint} of the Planck and BICEP2/Keck Array data, there is an upper limit on the ratio $r$ between the ratio $r$ of the dimensionless power spectra of gravitational waves to scalar perturbations, namely $r < 0.12$ (at a $2\sigma$ confidence level). Future experiments will be able to probe the possible presence of primordial gravitational waves on cosmlogical scales to a fainter level \cite{Whitebook}. From the theoretical perspective, a careful characterization of the power spectrum of B-mode polarization is necessary in order to distinguish between the predictions of different paradigms of very early universe cosmology. The most prevailing paradigm for early universe cosmology is the inflationary scenario, in which both the scalar and tensor primordial perturbations are predicted to be roughly scale invariant with slightly red tilts of which the values are determined by ``slow roll'' parameters and which satisfy the Planck data very well (see e.g. \cite{Vennin}). In the simplest realizations of single field slow roll inflation there is in fact a consistency relation between the tilt $n_t$ of the tensor spectrum and the value of $r$, namely $ n_t = -r/8 $. In particular, this implies that the tensor tilt is negative. In fact, in the context of slow roll inflation with standard vacuum initial conditions a negative tilt can only be avoided if matter which violates the Null Energy Condition is introduced. However, several theoretically sound mechanisms were put forward in the literature to extend the standard inflationary paradigm of single field slow roll so that a blue tensor tilt could become possible \cite{Wang:2014kqa} , such as Galileon inflation to be discussed below,  the beyond-slow-roll inflation \cite{Gong:2004kd, Gong:2014qga} , the non-Bunch-Davies inflation \cite{Easther:2001fi, Chen:2006nt, Ashoorioon:2014nta, Jiang:2015hfa, Jiang:2016nok} , the non-commutative (field) inflation \cite{Alexander:2001dr}.

Another successful scenario of the very early universe, whose idea originated from the cosmological application of superstring theory, is String Gas Cosmology (SGC) \cite{Brandenberger:1988aj} . In this scenario, a nearly scale invariant spectrum of primordial density fluctuations with a red tilt can be achieved from thermal fluctuations of the string gas satisfying a holographic heat capacity condition \cite{Nayeri:2005ck, Brandenberger:2006xi, Brandenberger:2006vv} . Interestingly, in this cosmological paradigm the power spectrum of primordial gravity waves is predicted to be slightly blue, and hence distinguishable from standard inflationary models. However, the blueness of the tensor spectrum would not remove the degeneracy with the above extended inflationary models mentioned in the previous paragraph. Fortunately, the SGC scenario predicts a specific consistency relation between the spectral index $n_s$ of the scalar perturbations and that of the tensor modes \cite{Brandenberger:2006xi} , which is given by
\begin{eqnarray}
 n_t \simeq 1-n_s ~,
\end{eqnarray}
at leading order. This SGC consistency relation differs from the predictions of the aforementioned inflationary models, and hence, if the spectral index of primordial tensor fluctuations can be measured in future CMB surveys, it may provide a smoking gun to distinguish the SGC scenario not only from standard inflation, but also from the inflation models mentioned above. Additionally, the primordial non-Gaussianities produced by the SGC are typically scale dependent and observationally invisible at current astronomical instruments \cite{Chen:2007js} .

Recently, however, another class of non-canonical inflation models were proposed in which ``higher-derivative" Horndeski operators (also called the Galilean operators) were introduced. The resulting scenario is called G-inflation \cite{Kobayashi:2010cm, Deffayet:2011gz, Kobayashi:2011nu} . This type of inflation model the NEC can be violated, and hence it is possible to generate a blue tilt for primordial tensor fluctuations and at the same time maintain a red tilt for scalar modes. Although it was shown in Ref. \cite{Cai:2014uka} via a phase space analysis that the background evolution to realize this scenario is not very natural  (the corresponding inflaton trajectory is not a fixed point), the G-inflation model can be degenerate with the SGC paradigm \footnote{By breaking local diffeomorphism invariance an inflationary model can be constructed which also leads to a blue spectrum of tensor modes \cite{Canone} which maintaining the red spectrum of scalar modes \cite{Leila}.}.

In order to break the degeneracy between the SGC and G-inflation scenarios, we in the present article utilize the ``effective field theory'' approach to inflation \cite{Cheung:2007st} to analyze under which conditions the model of G-inflation can reproduce the specific consistency relation of SGC. We further calculate the corresponding non-Gaussianities of G-inflation and show that they are observationally large unless an extremely unnatural fine tuning is performed to suppress these nonlinear fluctuations. Even so, the primordial non-Gaussianities generated in G-inflation are almost scale-independent in the squeezed limit, and hence, these predictions are different from those of SGC. Therefore, in principle one could distinguish these two very early universe scenarios by measuring the scale dependence of the primordial non-Gaussianities hidden in the CMB temperature anisotropy maps.

Another method to distinguish SGC and inflationary scenarios, including G-inflation is through the interference between the density fluctuation and additional massive fields \cite{Chen:2009we, Chen:2009zp, Baumann:2011nk, Chen:2012ge} . The method is known as quantum primordial standard clocks \cite{Chen:2015lza, Chen:2016cbe} because the oscillatory phase of the massive fields behaves as a clock to record the time and thus evolution history of the primordial universe. The conclusions in \cite{Chen:2015lza, Chen:2016cbe} directly apply to our study. Thus we shall not discuss this approach in detail in this work.

This article is organized as follows. In Section 2, we briefly review the paradigm of SGC and show how it produces a blue tilted power spectrum for primordial tensor modes as well as scale-dependent and observationally small non-Gaussianities. Then, in Section 3, we study cosmological perturbation theory within the inflationary paradigm based on the effective field approach. In particular, we focus on two topics. One is to analyze the conditions how to produce a SGC-like consistency relation within G-inflation; and the other is to study general properties of primordial non-Gaussianities of G-inflation and their comparison with those of SGC. Afterwards, we summarize in Section 4 the methods of distinguishing the G-inflation model from the SGC scenario with concluding remarks. The calculations of the non-Gaussianities are given in the Appendix.

\section{The SGC: Emergent Stringy Universe}

The SGC scenario, which describes the evolution of a gas of fundamental strings in the early universe based on the T-duality symmetry, was initially proposed in \cite{Brandenberger:1988aj} (see also \cite{Perlt}) to explain why only three of the nine microscopic spatial dimensions of string theory eventually becomes macroscopic. Based on this paradigm, a theory of cosmological perturbations was then later developed \cite{Nayeri:2005ck} under the assumption that the perturbed Einstein equations on cosmological scales are valid from the very initial Hagedorn phase of string gas matter until the final thermal equilibrium of regular radiation matter.

\begin{figure}[tbp]
\centering 
\includegraphics[width=.45\textwidth]{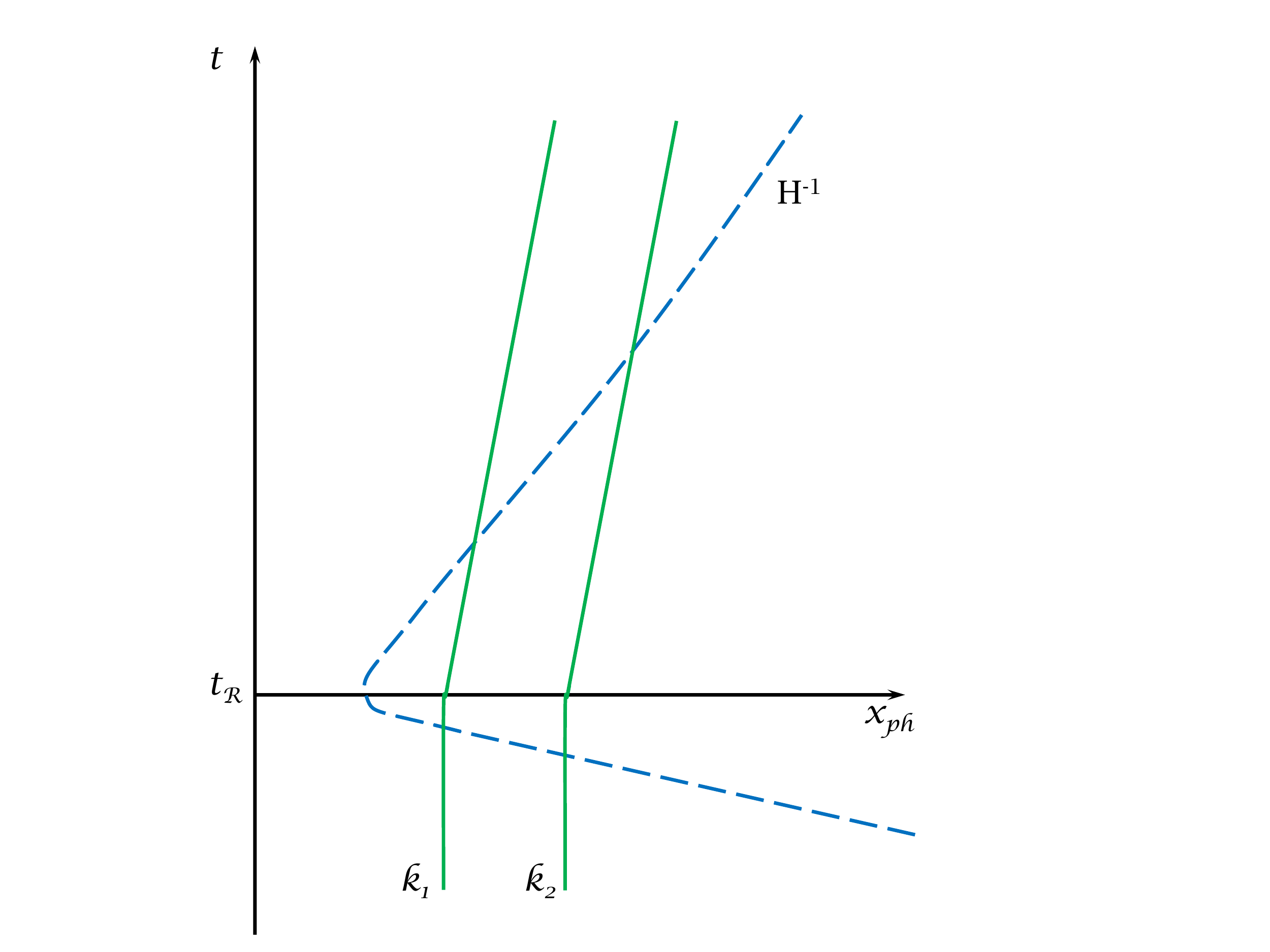}
\caption{\label{Fig:sketch}
A sketch of the evolution of two Fourier modes (labeled by $k_1$ and $k_2$) of cosmological perturbations in SGC. The vertical and horizontal axes are the physical time $t$ and space distance $x_{ph}$. The moment $t_R$ denotes when the universe transits from the stringy Hagedorn phase to the regular radiation dominated phase. The blue dashed curve represents the physical Hubble radius $H^{-1}$ and the green solid curves denote the wavelengths of two Fourier modes of cosmological perturbation labeled by comoving wave numbers $k_1$ and $k_2$.
}
\end{figure}

According to the SGC paradigm, one can study the perturbation theory by expanding the metric to linear order and then tracking the Fourier modes of the metric fluctuation one by one, such as sketched in Fig.~\ref{Fig:sketch}. In this figure, the blue dashed line represents the Hubble radius $H^{-1}$, which was infinity during the Hagedorn phase but quickly shrinks to a microscopic minimal length at the end of the early string gas phase, which corresponds to the reheating moment $t_R$ in inflationary cosmology \footnote{In SGC no separate reheating mechanism is required since the decay of string winding modes to string loops at the end of the Hagedorn phase directly produces a hot radiation bath.}, and then transitions to the regular thermal expansion as in Standard Big Bang cosmology. As shown in the sketch, primordial fluctuations, for instance the Fourier modes denoted by $k_1$ and $k_2$, were able to exit the Hubble radius due to the quick decrease of the Hubble radius without changing their physical wavelengths. Since fluctuations begin on sub-Hubble scales, a causal generation mechanism is possible, and the natural assumption is that the fluctuations are classical thermal fluctuations of a string gas.

In string theory, the internal energy of a system with temparature $T$ close to the string Hagedorn temperature $T_H$ is given by
\begin{align}
  \label{eq:hagedorn}
  \left \langle U \right \rangle \simeq
  \frac{L^2}{l_s^3} \ln \left [ \frac{l_s^3 T}{L^2 (1-T/T_H) } \right ]~,
\end{align}
where $L$ is the size of the system, and  $l_s$ is the string length. Assuming that matter is a gas of fundamental closed strings on a compact space at a temperature $T$ close to the Hagedorn temperature, then it follows from the partition function of a string gas \cite{Tan} that the root mean square of the density fluctuation $\delta \rho =\rho - \langle \rho \rangle$ in a volume V of length L is given by
\begin{align}
  \label{eq:sgc2pt}
  \left \langle \delta\rho^2 \right \rangle
  = \frac{\left \langle \delta U^2 \right \rangle}{V^2}
  = - \frac{1}{V^2} \frac{d \langle U\rangle}{d\beta} = \frac{T}{L^4 l_s^3 (1-T/T_H)}~,
\end{align}
where $\beta \equiv 1/T$. To calculate the power spectrum in momentum space, note that $\delta\rho_\mathbf{k} \sim k^{-3/2}\delta\rho$. The Fourier mode $\rho_ \mathbf{k}$ can then be related to the fluctuation of the gravitational potential using the Poisson equation
\begin{align}
  \label{eq:poisson}
  \Phi_ \mathbf{k} \sim 4\pi G \delta\rho_ \mathbf{k} \left ( \frac{a}{k}  \right )^2~,
\end{align}
where the gravitational potential $\Phi_ \mathbf{k}$ follows from the metric fluctuation in Newtonian gauge
\begin{align}
  \label{eq:newtonian-gauge}
  ds^2 = a^2 \left [ -(1+2\Phi) d\tau^2 + (1-2\Phi) d \mathbf{x}^2 \right ]~.
\end{align}
The power spectrum of the gravitational potential can thus be calculated to be
\begin{align}
  \label{eq:pPhi}
  P_\Phi (k)= 8 \left ( \frac{l_p}{l_s}  \right )^4 \frac{1}{1-T(k)/T_H} ~,
\end{align}
where $l_p$ is the Planck length and $T(k)$ is the temperature of the string gas when the scale $k$ exits the Hubble radius towards the end of the Hagedorn phase, and we have ${l_{p}}=\frac{1}{\sqrt{8\pi }}$ in units where $M_p=1$ (in the following we will set $M_p=1$). Note that the temperature slowly decreases while the universe departs from the Hagedorn phase. Since $P_\Phi (k)$ is an increasing function of $T(k)$, it is a decreasing function of $k$ (recall that smaller $k$ modes exit Hubble radius earlier, at a higher temperature). As a result, a red tilt of the power spectrum of scalar cosmological perturbations is predicted \cite{Nayeri:2005ck}.

The non-Gaussianity of String Gas Cosmology can be similarly calculated \cite{Chen:2007js}. By noting that in a thermal system equation (where in the final step we have used the same assumptions which went into \eqref{eq:sgc2pt})
\begin{align}
  \label{eq:sgc3pt}
  \left \langle \delta\rho^3 \right \rangle
  = \frac{\left \langle \delta U^3 \right \rangle}{V^3}
  = - \frac{1}{V^3} \frac{d^2 \langle U\rangle}{d\beta^2} = \frac{T^2}{L^7 l_s^3 (1-T/T_H)^2}~.
\end{align}
one can relate $\delta\rho$ to $\Phi_ \mathbf{k}$, and one obtains the parameter $f_{NL}$
\begin{align}
  \label{eq:fnl}
  f_{NL} \sim k^{-\frac{3}{2} } \frac{\langle\Phi_ \mathbf{k}^3\rangle}{\langle\Phi_ \mathbf{k}^2\rangle^2}
  \sim \frac{l_s^3 H(t_H(k))}{4\pi l_p^2} \sim \left ( \frac{l_s}{l_p}  \right )^2 \frac{H_0}{T_0} \frac{k}{k_0} \sim 10^{-30} \left ( \frac{l_s}{l_p}  \right )^2 \frac{k}{k_0}  ~,
\end{align}
where $k_0$ corresponds to the scale which returns to the horizon at the present time. The subscript 0 means the values measured today. $H_0 \sim 10^{-18} s^{-1}$ and $T_0 \sim 2.7 K$. It is easy to restore the dimension by multiplying $\hbar/k_B$ where $\hbar$ is the reduced Planck constant and $k_B$ is the Boltzmann constant. Two conclusions can be drawn from \eqref{eq:fnl}. The fluctuations on cosmological scales in String Gas Cosmology are highly Gaussian, unless the string scale is around TeV scale (which is very unlikely considering that the LHC has not observed a stringy signature). Note also that the non-Gaussianities of String Gas Cosmology are strongly scale dependent.

The tensor fluctuations of String Gas Cosmology can be calculated \cite{Brandenberger:2006xi} by evaluating the fluctuations of the spatial part of the stress tensor $\delta T_{ij}$. From the space-space part of the Einstein equations,
\begin{align}
  \label{eq:space-einstein}
  k^2 h_{ij}(k) \sim 8\pi G \delta T_{ij}(k)~,
\end{align}
and the power spectrum of gravitational waves $h_{ij}$ can be calculated by computing the expectation value $\langle \delta T_{ij} \delta T^{ij} \rangle$ \cite{Chen:2007js, Brandenberger:2014faa} ,
\begin{align}
  \label{eq:ph}
  P_h \sim \left ( \frac{l_p}{l_s}  \right )^4 \left ( 1 - \frac{T}{T_H}  \right ) \ln^2 \left [ \frac{1}{l_s^2 k^2} \left ( 1-\frac{T}{T_H}  \right )  \right ]~.
\end{align}
Up to logarithmic corrections, the tensor spectrum depends linearly on $1-T/T_H$, instead of $1/(1-T/T_H)$ as in the scalar case. Thus the tensor spectrum has a blue tilt. This prediction distinguishes SGC compared to most inflation models.

\section{Effective Field Theory of Inflation and Non-Gaussianities}

To model-independently investigate the predictions of Galileon inflation scenarios, we use the framework of effective field theory of inflation \cite{Cheung:2007st}. The effective field theory of inflation is a way to treat perturbations systematically if we do not care about the evolution of the background solution $\phi_0(t)$. This framework is versatile enough in the sense that it can describe all possible single field models, like DBI inflation which corresponds to its large $(g^{00}+1)^n$ limit and ghost inflation which corresponds to its $\dot{H}\rightarrow 0$ limit. We use this framework to discuss possible consequences of the blue tilted tensor power spectrum on inflation because, in this language, we can directly see the constraints on parameter space if we require the tensor spectrum to be blue tilted.

First we separate the inflaton into background and perturbations, $\phi(\mathbf x,t)=\phi(t)+\delta \phi(\mathbf x,t)$. Here $\phi(\mathbf x,t)$ is a scalar under all differeomorphisms. $\phi(t)$ and $\delta \phi(\mathbf x,t)$ are invariant only under spatial differeomorphisms. Under a time differeomorphism $t\rightarrow t^\prime = t+\xi^0(t,\mathbf x)$, the background solution $\phi(t)$ transforms as
\begin{equation} \label{timediff}
\phi (t)\rightarrow \phi(t)-\dot{\phi}(t)\xi^0~,
\end{equation}
thus leading to the following transformation for the perturbation:
\begin{equation}
\delta\phi (\mathbf x,t)\rightarrow \delta\phi(\mathbf x,t)+\dot{\phi}(t)\xi^0~.
\end{equation}
We choose the inflaton flat gauge $\delta \phi=0$. In this gauge, the inflaton is eaten by the graviton and leads to three degrees of freedom, namely two tensor modes and one scalar mode.

As described in \cite{Cheung:2007st} , the most general Lagrangian which respects time dependent spatial differeomorphisms but can break time differeomorphism invariance can be written as
\begin{equation}
S=\int d^4x \sqrt{-g} F(R_{\mu\nu\rho\sigma},g^{00},K_{\mu\nu},\nabla_\mu,t)~,
\end{equation}
where $R_{\mu\nu\rho\sigma}$ is the Riemann tensor which respects all differeomorphisms, $g^{00}$ is the time component of the metric and $K_{\mu\nu}$ is the extrinsic curvature.
The extrinsic curvature is
\begin{equation}
K_{\mu\nu}=(\delta^\sigma_\mu+n^\sigma n_\mu)\nabla_\sigma n_\nu~,
\end{equation}
where $n_\mu$ is the unit vector perpendicular to the hypersurface
\begin{equation}
n_\mu=(-N,0,0,0)=-\delta^0_\mu/\sqrt{-g^{00}},\quad n^\nu=(1/N,N^i/N)~,
\end{equation}
where $N$ and $N_i$ are defined through the ADM decomposition
\begin{equation}
ds^2=-N^2dt^2+h_{ij}(N^idt+dx^i)(N^jdt+dx^j)~,
\end{equation}

The most general action with broken time differeomorphism symmetry but preserving time dependent spatial differeomorphisms about a given FRW background is
\begin{align} \label{eq:eft-df-mix}
  S = & \int d^4x \sqrt{-g} [
    \frac{1}{2} R + \dot{H}g^{00}-(3H^2 +\dot{H})+\frac{1}{2!}M_2^4(g^{00}+1)^2
    \nonumber\\ &
    + \frac{1}{3!} M_3^4 (g^{00}+1)^3 -\frac{1}{2} \bar{M}_1^3 (g^{00}+1)\delta K^\mu{}_\mu-\frac{1}{2}\bar{M}_2^2\delta K^\mu{}_\mu{}^2-\frac{1}{2}\bar{M}_3^2\delta K^\mu{}_\nu \delta K^\nu{}_\mu
    \nonumber\\ &
    -\frac{1}{3!} \bar{M}_4^3 (g^{00}+1)^2 \delta K^\mu{}_\mu-\frac{1}{3!}\bar{M}_5^2(g^{00}+1)\delta K^\mu{}_\mu{}^2-\frac{1}{3!}\bar{M}_6^2(g^{00}+1)\delta K^\mu{}_\nu \delta K^\nu{}_\mu
    \nonumber\\ &
    -\frac{1}{3!}\bar{M}_7\delta K^\mu{}_\mu{}^3-\frac{1}{3!}\bar{M}_8\delta K^\mu{}_\mu \delta K^\nu{}_\rho \delta K^\rho{}_\nu-\frac{1}{3!} \bar{M}_9\delta K^\mu{}_\nu \delta K^\nu{}_\rho \delta K^\rho{}_\mu
  ]~,
\end{align}
where $\delta K_{\mu\nu}$ is the variation of the extrinsic curvature $\delta K_{\mu\nu} = K_{\mu\nu}-H h_{\mu\nu}$. All coefficientds in this action can have a time dependence. However we are only interested in one Hubble time step, during when $H$, $\dot{H}$ term and the other operators do not vary too much.

In the effective field theory of inflation, one can perform a time-diffeomorphism with $\xi^0$ (see above (\ref{timediff})), and relate the parameter to a fluctuation field $\pi$ which changes under time diffeomorphisms: $\pi(t,\mathbf{x})\to\pi(t,\mathbf{x})-\xi^0(t,\mathbf{x})$. This fluctuating scalar field $\pi$ is the Goldstone boson for time diffeomorphism symmetry breaking.

To be specific, under a time diffeomorphism we have
\begin{align}
&g^{ij}\rightarrow g^{ij}~~~,~~~ g^{0i}\rightarrow(1+\dot{\pi})g^{0i}+\partial_j\pi g^{ji}~,\nonumber\\
&g^{00}\rightarrow(1+\dot{\pi})^2 g^{00}+2(1+\dot{\pi})g^{0i}\partial_i\pi+g^{ij}\partial_i\pi\partial_j\pi
=-(1+\dot{\pi})^2 +\frac{(\partial_i\pi)^2}{a^2}~.
\end{align}
Thus, the variation of the extrinsic curvature to third order is
\begin{align}
&\delta K_{00}=0+\mathcal{O}(\pi^3)~~~,~~~\delta K_{0i}=-\frac{\partial_j\pi(\partial_j\partial_i\pi)}{a^2}+\mathcal{O}(\pi^3)~,\nonumber\\
&\delta K_{ij}=-\partial_i\partial_j\pi-H\partial_i\pi\partial_j\pi+\partial_i\dot{\pi}\partial_j\pi+\partial_j\dot\pi\partial_i\pi +\dot\pi\partial_i\partial_j\pi+\frac{1}{2}(\partial_a\pi)^2\delta_{ij}H+\mathcal{O}(\pi^3)~,
\end{align}
which allows us to expand the action from the fluctuation field $\pi$ keeping second and third order terms. This yields
\begin{equation}\label{secondLag}
{{S}_{2}}=\int{{{d}^{4}}}x[-\frac{1}{H\tau }(2M_{2}^{4}-\dot{H})\pi {{'}^{2}}-\frac{1}{H\tau }(\dot{H}+\frac{\bar{M}_{1}^{3}H}{2}){{({{\partial }_{i}}\pi )}^{2}}+(\frac{\bar{M}_{2}^{2}}{2}+\frac{\bar{M}_{3}^{2}}{2})H\tau {{({{\partial }_{i}}{{\partial }_{j}}\pi )}^{2}}]~,
\end{equation}
\begin{align}
  & {{S}_{3}}=\int{{{d}^{4}}}x[(2M_{2}^{4}-\frac{4M_{3}^{4}}{3}){{{{\pi }'}}^{3}}-(2M_{2}^{4}+\frac{\bar{M}_{1}^{3}H}{2}){\pi }'{{({{\partial }_{i}}\pi )}^{2}}+H\tau \bar{M}_{1}^{3}{\pi }'{{\partial }_{i}}\pi {{\partial }_{i}}{\pi }' \nonumber\\
 & +(\frac{\bar{M}_{1}^{3}}{2}-\frac{\bar{M}_{2}^{2}H}{2}-\bar{M}_{3}^{2}H)H\tau {{({{\partial }_{i}}\pi )}^{2}}{{\partial }_{j}}{{\partial }_{j}}\pi +\bar{M}_{2}^{2}{{(H\tau )}^{2}}{{\partial }_{i}}\pi {{\partial }_{i}}{\pi }'{{\partial }_{j}}{{\partial }_{j}}\pi -\bar{M}_{2}^{2}{{(H\tau )}^{2}}{\pi }'{{\partial }_{i}}\pi {{\partial }_{i}}{{\partial }_{j}}{{\partial }_{j}}\pi   \nonumber\\
 & +(\bar{M}_{3}^{2}+\frac{\bar{M}_{6}^{2}}{3}){{(H\tau )}^{2}}{\pi }'{{({{\partial }_{i}}{{\partial }_{j}}\pi )}^{2}}-\bar{M}_{3}^{2}{{(H\tau )}^{2}}{{({{\partial }_{i}}\pi )}^{2}}{{\partial }_{j}}{{\partial }_{j}}{\pi }'+\frac{2\bar{M}_{4}^{3}}{3}H\tau {{{{\pi }'}}^{2}}{{\partial }_{i}}{{\partial }_{i}}\pi +\frac{\bar{M}_{5}^{2}}{3}{{(H\tau )}^{2}}{\pi }'{{({{\partial }_{i}}{{\partial }_{i}}\pi )}^{2}}  \nonumber\\
 & +\frac{{{{\bar{M}}}_{7}}}{6}{{(H\tau )}^{3}}{{({{\partial }_{i}}{{\partial }_{i}}\pi )}^{3}}+\frac{{{{\bar{M}}}_{8}}}{6}{{(H\tau )}^{3}}{{\partial }_{i}}{{\partial }_{i}}\pi {{({{\partial }_{j}}{{\partial }_{k}}\pi )}^{2}}+\frac{{{{\bar{M}}}_{9}}}{6}{{(H\tau )}^{3}}{{\partial }_{i}}{{\partial }_{j}}\pi {{\partial }_{j}}{{\partial }_{k}}\pi {{\partial }_{k}}{{\partial }_{i}}\pi ]~,
\end{align}
where a prime means a derivative with respect to conformal time $\tau$.

It was believed for a long time that if the background solution violates the NEC, then there will be a ghost in the inflationary scalar sector. However, it was noticed later that if we have a Galileon term in the inflation sector, the NEC can be violated without encountering a ghost. In this case, a slightly blue tensor tilt $n_t$ can  be generated. An example is the G-inflation model of \cite{Cai:2014uka} given by the action\footnote{The action is considered in \cite{Cedric:2010kgb} for constructing models of dark energy.}
\begin{equation}
S=\int{{{d}^{4}}}x\sqrt{-g}\left( K+G\square \phi  \right) ~,
\end{equation}
where
\begin{equation}
K=-X-V~~~,~~~G=-2\gamma X ~,
\end{equation}
and $X=-\frac{1}{2}\nabla^\mu\phi\nabla_\mu\phi$. Here, $V$ is the slow-roll potential and $\gamma$ is a parameter. In the Goldstone language, the corresponding effective field theory can be written in terms of the action \cite{Burrage:2010cu}
\begin{equation}
S=\int{{{d}^{4}}}x{{a}^{3}}\left[ Q_1 ({{{\dot{\pi }}}^{2}}-\frac{c_{s}^{2}}{{{a}^{2}}}{{({{\partial }_{i}}\pi )}^{2}})+{{\omega }_{1}}{{{\dot{\pi }}}^{3}}+\frac{{{\omega }_{3}}}{{{a}^{2}}}\dot{\pi }{{({{\partial }_{i}}\pi )}^{2}}+\frac{{{\omega }_{4}}}{{{a}^{4}}}{{({{\partial }_{i}}\pi )}^{2}}({{\partial }_{j}}{{\partial }_{j}}\pi ) \right]~,
\end{equation}
where
\begin{align}
  & {{c}_{2}}=1~~~,~~~{{c}_{3}}=\gamma~~~,~~~Z=H\dot{\phi }~,  \nonumber\\
 & {{Q}_{1}}=\frac{{{{\dot{\phi }}}^{2}}}{2}({{c}_{2}}+12{{c}_{3}}Z)~~~,~~~{{Q}_{2}}=\frac{{{{\dot{\phi }}}^{2}}}{2}({{c}_{2}}+4{{c}_{3}}(2Z+\ddot{\phi }))~, \nonumber\\
 & {{\varsigma }_{1}}=2H{{{\dot{\phi }}}^{3}}{{c}_{3}}~~~,~~~{{\varsigma }_{2}}=-2{{{\dot{\phi }}}^{3}}{{c}_{3}}~~~,~~~{{\varsigma }_{3}}=(-2H{{{\dot{\phi }}}^{3}}+2{{{\dot{\phi }}}^{2}}\ddot{\phi }){{c}_{3}}~~~,~~~{{\varsigma }_{4}}={{{\dot{\phi }}}^{3}}{{c}_{3}}~,  \nonumber\\
 & c_{s}^{2}=\frac{{{Q}_{2}}}{{{Q}_{1}}}~~~,~~~{{\omega }_{1}}={{\varsigma }_{1}}+\frac{2H}{c_{s}^{2}}{{\varsigma }_{2}}+\frac{2}{3c_{s}^{2}}\frac{{{{\dot{Q}}}_{1}}}{{{Q}_{1}}}-\frac{{{Q}_{2}}}{3c_{s}^{2}}\frac{d}{dt}\left( \frac{{{\varsigma }_{2}}}{{{Q}_{2}}} \right)
~~~,~~~{{\omega }_{3,4}}={{\varsigma }_{3,4}}~.
\end{align}
For this model, a detailed dynamical system analysis \cite{Cai:2014uka} shows that it is possible to obtain a red tilted scalar perturbation and a blue tilt for tensor perturbations while protecting the de Sitter background solution at the same time. This analysis shows us that it is impossible to distinguish String Gas Cosmology from G-inflation based only on the signs of spectral indices.

Here, we point out that non-Gaussianity can be a signal to distinguish String Gas Cosmology from inflationary models with a blue tensor tilt. Inflationary models with a blue tilt of the tensor spectrum generically produce large non-Gaussianities, whereas the predicted non-Gaussianities from String Gas Cosmology are negligible on cosmological scales \footnote{We are assuming here that no stable cosmic superstrings \cite{Witten, CMP} survive as a remnant of the early Hagedorn phase to late times. If such cosmic superstrings were stable, then there would be specific non-Gaussian features in position space maps (see e.g. \cite{RHBCSrev}) which once again are easy to distinguish from primordial inflationary non-Gaussianities.}. We will demonstrate this point using the effective field theory approach to inflation. The advantage is that, when using effective field theory we can treat the models in a general way. The analysis will include all inflation models with Galileons. The non-Gaussianities in the effective field theory of inflation have been computed in previous works (for example \cite{Bartolo:2010bj} ), but their results are numerical and the studies did not explore the parameter space yielding a blue tensor spectrum.

From the second order Lagrangian we can read the condition for the absence of ghosts:
\begin{equation}
2M^4_2-\dot{H}>0~.
\end{equation}
We also have to worry about gradient instabilities. Two possibilities to avoid them can be seen from (\ref{secondLag}): If we require the $(\partial_i\pi)^2$ terms have negative coefficients, then we require
\begin{equation}
\frac{\bar{M}^3_1H}{2}+\dot{H}<0~.
\end{equation}
If we loosen this constraint and take the $(\partial_i\partial_j \pi)^2$ terms, then we obtain a less strict condition
\begin{equation}
\frac{\dot{H}}{H^2}+\frac{a^3\bar{M}^3_1}{2H}-(\bar{M}^2_2+\bar{M}^2_3)\frac{a^3}{2}<0~.
\end{equation}

We have calculated the shape of non-Gaussianities (as measured by the three point function) and summarize the analysis in the Appendix, where we also identify a special regime of parameters for which the theory can yield small non-Gaussianities. Applied to the present model we can calculate the contributions to $f_{NL}$ from the $M_2$ and $\bar{M}_1$ terms, and estimate their order of magnitude taking into account the inequalities in the second case above. According to \cite{Chen:2010xka} , we obtain
\begin{equation}
\langle\pi^3\rangle=\frac{-1}{H^3}\langle\zeta^3\rangle=\frac{-(2\pi)^7}{H^3}\delta^{(3)}(\mathbf{k}_1+\mathbf{k}_2+\mathbf{k}_3)\frac{P^2_{\zeta}}{(k_1k_2k_3)^2}S(k_1,k_2,k_3)~.
\end{equation}
In the $k_1=k_2=k_3$ limit
\begin{equation}
S(k_1,k_2,k_3)\rightarrow \frac{9}{10}f_{NL} ~,
\end{equation}
Finally, combining the constraints on the parameters, we obtain
\begin{equation}
f_{NL(M_2)}>\frac{35}{108}\sim \mathcal{O}(1),~f_{NL(\bar{M}_1)}<-\frac{85}{108}\sim \mathcal{O}(1)~,
\end{equation}
where $f_{NL(M_2)}$ is the non-Gaussianity produced by allowing only $M_2$ to be non-zero, and $f_{NL(\bar{M}_1)}$ is the non-Gaussinity produced by allowing only $\bar{M}_1$ to be non-zero. As we expect, we obtain large non-Gaussianities outside the regime that is given in the Appendix, for which region the non-Gaussianities may be canceled by fine-tuning.

In the effective field theory of inflation, the scalar sector is more general than the minimal inflation model. Although the modification is more transparent on the scalar field, sometimes it can also be considered as modified gravity. This is because of two reasons: On the one hand, the scalar degree of freedom can be eaten by the gravity sector via a gauge transformation. On the other hand, for generalized G-inflation, the Lagrangian contains non-trivial dependence on the metric and even non-trivial dependence on the Ricci curvature because of the high derivative terms on the scalar field, and thus gravity is modified. All those modifications are included in the effective field theory approach to inflation.

\section{Methods to Distinguish String Gas Cosmology from Inflationary Cosmologies}

Since String Gas Cosmology predicts a blue tilt of the tensor spectrum, a careful measurement of the tensor tilt (see e.g. \cite{tensorconstraints} for current and predicted observational limits) can be used to distinguish String Gas Cosmology from simple scalar field models (with actions based on a canonical kinetic term) which always predict a red tilt. However, there are inflationary models which result from abandoning some of the physical principles of standard inflation which predict a blue tensor tilt.

In the previous sections we have explored some ways of distinguishing String Gas Cosmology from these non-standard inflation models. The first is based on the predicted non-Gaussianities as measured by the three point correlation function. Because of the no-ghost constraint inflationary models with a blue tensor tilt usually leads to non-Gaussianities of the order of at least $\mathcal{O}(1)$ in the $f_{NL}$ parameter, when computed at lowest order in the effective field theory expansion. If we take higher order corrections into account, fine-tuning for some special shapes in parameter space (see Appendix) can lead to a cancellation between large terms in the expressions for the three point function, even providing a vanishing $f_{NL}$. Therefore, this doubly tuned class of inflation models (first tuning meaning non-simple standard canonical slow-roll inflation, second tuning being the tuning in parameter space) cannot be differentiated from String Gas Cosmology even my means of the smallness of the three point function. But now a further consideration comes into play: the scale dependence of non-Gaussianities. Considering that the small non-Gaussianities generated by inflation are scale independent, while String Gas Cosmology has a scale dependent $f_{NL}$, we can in principle also distinguish String Gas Cosmology from inflationary cosmologies even for small values of $f_{NL}$.

One can also distinguish these two cosmologies through the consistency relation. As discussed earlier, String Gas Cosmology leads to a special consistency relation between the tensor and scalar spectral indices, namely
\begin{equation}
n_s=1-n_t \, .
\end{equation}
This is different from the consistency relations in inflationary models.

For inflation, first let us take a look at the scalar spectrum. In the effective field theory description of inflation, for a special limit $\bar{M}_1=\bar{M}_2=\bar{M}_3=0$ we get the DBI-type relation as (see \cite{Bartolo:2010im})
written in terms of the inflationary slow-roll parameters $\epsilon$ and $\eta$
\begin{align}
n_s-1=-2\epsilon-\eta+f_1(\epsilon,\eta,...) ~,
\end{align}
where $f_1$ represents higher order terms in the slow-roll expansion (and $f_2$ in the following is also such a function), and is related to coefficients in the effective field theory of inflation. For other cases the relation is given as
\begin{align}
n_s-1=-4\epsilon+f_2(\epsilon,\eta,...) ~.
\end{align}
These relations hold in the framework of the effective field theory description of inflation for minimally coupled gravity, which has the standard relation
\begin{equation}
n_t=-2\epsilon \, .
\end{equation}
As a result, if we do not abandon the slow roll expansion and minimally coupled gravity it is extremely hard to reproduce the consistency relation in String Gas Cosmology (except for fine tunings including higher order corrections in slow roll expansion), which means that $n_s=1-n_t$ can be used as a means to distinguish String Gas Cosmology from inflationary models. For G-inflation in modified gravity these formulas may be different, but fine-tuning will also be needed to reproduce a string gas consistency relation.

To summarize, we have presented several methods to distinguish String Gas Cosmology from inflationary models with a blule tensor spectrum:
\begin{itemize}
\item String Gas Cosmology is ruled out if $f_{NL}$ is at or larger than $\mathcal{O}(1)$ \footnote{Except if the non-Gaussianities have the special geometrical patterns predicted by cosmic strings and cosmic superstrings.}.
\end{itemize}
\begin{itemize}
\item String Gas Cosmology and some inflationary models both provide a small three point function, but $f_{NL}$ is scale independent for inflation, but scale dependent for String Gas Cosmology.
\end{itemize}
\begin{itemize}
\item String Gas Cosmology has a special consistency relation $n_s=1-n_t$, which is extremely hard to produce from inflation.
\end{itemize}

\section{Conclusions and Outlook}

In the present article, we explicitly show that we can distinguish String Gas Cosmology from the class of special inflation models which can produce blue tilted tensor power spectrum. The first approach is by using properties of the three point function. In this context there are two kinds of ways to distinguish the two classes of models; one is through the magnitude of non-Gaussianities, the other is by whether the non-Gaussianities are scale dependent or not. We find that only inflationary models can produce large non-Gaussianities. We show this point by explicitly calculating the predictions for the three point function in inflationary models using the effective field theory approach. We found that in order to satisfy the ghost free condition and gradient stability condition, the coefficients in the effective field theory have to satisfy a set of inequalities. These constraints on the coefficients will lead to terms in the effective field theory Lagrangian which generate large non-Gaussianities. Hence, if future observations yield both a blue tilted tensor power spectrum and large non-Gaussianities, we then know that the fluctuations must come from an inflationary model. However, for a special range of model parameters it is possible to have cancellation between individually large terms, leading to a small three point function. If the non-Gaussianities are small, then, in principle, one could still distinguish between String Gas Cosmology and the special class of inflation models according to whether the non-Gaussianities are scale dependent or not. For the String Gas Cosmology model, the non-Gaussianities being produced are scale dependent, whereas for the inflation models, the non-Gaussianities are scale independent. This criterion, however, assumes we can actually measure the small non-Gaussianities.

A second way to distinguish between String Gas Cosmology and inflationary models makes use of the consistency relations between the spectral indices.  String Gas Cosmology has a special consistency relation $n_s=1-n_t$, which is extremely hard to produce from inflation models. This is also an effective way to distinguish these two cosmologies.

\section*{Acknowledgement}
\noindent

We thank E. Ferreira, D. G. Wang and Z. Wang for useful discussions. We also thank Y. Chen and T. Wrase for helpful discussions about string phenomenology at the SPSC2016 conference. The research of RB is supported in part by funds from NSERC and the Canada Research Chair program. YFC and MH are supported in part by the Chinese National Youth Thousand Talents Program (Grant No. KJ2030220006), by the USTC start-up funding (Grant No. KY2030000049) and by the National Natural Science Foundation of China (Grant No. 11421303). The works of JL and SL are supported in part by the Fund for Fostering Talents in Basic Science of the National Natural Science Foundation of China (Grant No. J1310021), the Outstanding Student International Exchange Funding Scheme and the Yan Ji-Ci Class in the University of Science and Technology of China. MH, JL are also grateful for PhD assistantships from the departments of physics of Tokyo University and Caltech respectively. YW is supported by Grant HKUST4/CRF/13G issued by the Research Grants Council (RGC) of Hong Kong. SZ is supported by the the Hong Kong PhD Fellowship Scheme (HKPFS) issued by the Research Grants Council (RGC) of Hong Kong. Part of numerical computations are operated on the computer cluster LINDA in the particle cosmology group at USTC.

\section*{Appendix: Shapes of the Non-Gaussianities}

We denote the \textit{n}th term in the Hamiltonian as $H_3^{(n)}$. These Hamiltonians come from the third order action and are written in terms of conformal time $\tau$ instead of physical time $t$. We have $S_3=\sum_j{\int d\tau H_3^{(j)}(\tau)}$. Then the corresponding three-point correlation functions of $H_3^{(n)}$s are given as follows (we have omitted the $(2\pi)^3\delta^{(3)}(\mathbf{k}_1+\mathbf{k}_2+\mathbf{k}_3)(\frac{-1}{2\sqrt{\epsilon}})^6$ factor in every result, and use $k=k_1+k_2+k_3$):
\begin{itemize}
\item $H_3^{(1)}=(\frac{2M_2^4}{H\tau}-\frac{4M_3^4}{3H\tau})\pi'^3$
\end{itemize}
\begin{equation}
2\operatorname{Im}\int^0_{-\infty}d\tau\langle0|\pi^3H_3^{(1)}(\tau)|0\rangle=\frac{48M^4_2-32M^4_3}{H}\frac{1}{k_1k_2k_3k^3}~.
\end{equation}
\begin{itemize}
\item $H_3^{(2)}=-(\frac{2M_2^4}{H\tau}+\frac{\bar{M}_1^3}{2\tau})\pi'(\partial_i\pi)^2$
\end{itemize}
\begin{equation}
2\operatorname{Im}\int^0_{-\infty}d\tau\langle0|\pi^3H_3^{(2)}(\tau)|0\rangle=(\frac{4M^4_2}{H}+\bar{M}^3_1)\frac{\cos\langle\mathbf{k}_2,\mathbf{k}_3\rangle}{k_1k^2_2k^2_3}(\frac{-1}{k}-\frac{k_2+k_3}{k^2}-\frac{2k_2k_3}{k^3})+\text{2 perm}.
\end{equation}
\begin{itemize}
\item $H_3^{(3)}=\bar{M}_1^3\pi'\partial_i\pi\partial_i\pi'$
\end{itemize}
\begin{equation}
2\operatorname{Im}\int^0_{-\infty}d\tau\langle0|\pi^3H_3^{(3)}(\tau)|0\rangle=-2\bar{M}^3_1\frac{\cos\langle\mathbf{k}_2,\mathbf{k}_3\rangle}{k_1k^2_2}(\frac{2}{k^3}+\frac{6k_2}{k^4})+\text{5 perm}.
\end{equation}
\begin{itemize}
\item $H_3^{(4)}=(\frac{\bar{M}_1^3}{2}-\frac{\bar{M}_2^2H}{2}-\bar{M}_3^2H)(\partial_i\pi)^2\partial_j\partial_j\pi$
\end{itemize}
\begin{displaymath}
2\operatorname{Im}\int^0_{-\infty}d\tau\langle0|\pi^3H_3^{(4)}(\tau)|0\rangle=(2\bar{M}^3_1-2\bar{M}^2_2H-4\bar{M}^2_3H)\frac{\cos\langle\mathbf{k}_1,\mathbf{k}_2\rangle}{k^2_1k^2_2k_3}
\end{displaymath}
\begin{equation}
\times(\frac{-2}{k}-\frac{2(k_1k_2+k_2k_3+k_1k_3)}{k^3}-\frac{6k_1k_2k_3}{k^4})+\text{2 perm}.
\end{equation}
\begin{itemize}
\item $H_3^{(5)}=\bar{M}_2^2H\tau\partial_i\pi\partial_i\pi'\partial_j\partial_j\pi$
\end{itemize}
\begin{equation}
2\operatorname{Im}\int^0_{-\infty}d\tau\langle0|\pi^3H_3^{(5)}(\tau)|0\rangle=\bar{M}^2_2H\frac{\cos\langle\mathbf{k}_1,\mathbf{k}_2\rangle}{k_1^2k_3}(\frac{2}{k^3}+\frac{6(k_1+k_3)}{k^4}+\frac{24k_1k_3}{k^5})+\text{5 perm}.
\end{equation}
\begin{itemize}
\item $H_3^{(6)}=-\bar{M}_2^2H\tau\pi'\partial_i\pi\partial_i\partial_j\partial_j\pi$
\end{itemize}
\begin{equation}
2\operatorname{Im}\int^0_{-\infty}d\tau\langle0|\pi^3H_3^{(6)}(\tau)|0\rangle=-\bar{M}^2_2H\frac{\cos\langle\mathbf{k}_2,\mathbf{k}_3\rangle}{k_1k_2^2}(\frac{2}{k^3}+\frac{6(k_2+k_3)}{k^4}+\frac{24k_2k_3}{k^5})+\text{5 perm}.
\end{equation}
\begin{itemize}
\item $H_3^{(7)}=(\bar{M}_3^2+\frac{\bar{M}_6^2}{3})H\tau\pi'(\partial_i\partial_j\pi)^2$
\end{itemize}
\begin{equation}
2\operatorname{Im}\int^0_{-\infty}d\tau\langle0|\pi^3H_3^{(7)}(\tau)|0\rangle=(4\bar{M}^2_3+\frac{4\bar{M}^2_6}{3})H\frac{\cos^2\langle\mathbf{k}_2,\mathbf{k}_3\rangle}{k_1k_2k_3}(\frac{2}{k^3}+\frac{6(k_2+k_3)}{k^4}+\frac{24k_2k_3}{k^5})+\text{2 perm}.
\end{equation}
\begin{itemize}
\item $H_3^{(8)}=-\bar{M}_3^2H\tau(\partial_i\pi)^2\partial_j\partial_j\pi'$
\end{itemize}
\begin{equation}
2\operatorname{Im}\int^0_{-\infty}d\tau\langle0|\pi^3H_3^{(8)}(\tau)|0\rangle=-4\bar{M}^2_3H\frac{k_3\cos\langle\mathbf{k}_1,\mathbf{k}_2\rangle}{k^2_1k^2_2}(\frac{2}{k^3}+\frac{6(k_1+k_2)}{k^4}+\frac{24k_1k_2}{k^5})+\text{2 perm}.
\end{equation}
\begin{itemize}
\item $H_3^{(9)}=\frac{2\bar{M}_4^3}{3}\pi'^2\partial_i\partial_i\pi$
\end{itemize}
\begin{equation}
2\operatorname{Im}\int^0_{-\infty}d\tau\langle0|\pi^3H_3^{(9)}(\tau)|0\rangle=-\frac{8\bar{M}^3_4}{3}\frac{1}{k_1k_2k_3}(\frac{2}{k^3}+\frac{6k_3}{k^4})+\text{2 perm}.
\end{equation}
\begin{itemize}
\item $H_3^{(10)}=\frac{\bar{M}_5^2H\tau}{3}\pi'(\partial_i\partial_i\pi)^2$
\end{itemize}
\begin{equation}
2\operatorname{Im}\int^0_{-\infty}d\tau\langle0|\pi^3H_3^{(10)}(\tau)|0\rangle=\frac{4\bar{M}^2_5}{3}H\frac{1}{k_1k_2k_3}(\frac{2}{k^3}+\frac{6(k_2+k_3)}{k^4}+\frac{24k_2k_3}{k^5})+\text{2 perm}.
\end{equation}
\begin{itemize}
\item $H_3^{(11)}=\frac{\bar{M}_7}{6}H^2\tau^2(\partial_i\partial_i\pi)^3$
\end{itemize}
\begin{equation}
2\operatorname{Im}\int^0_{-\infty}d\tau\langle0|\pi^3H_3^{(11)}(\tau)|0\rangle=-2\bar{M}_7H^2\frac{1}{k_1k_2k_3}(\frac{8}{k^3}+\frac{24(k_1k_2+k_2k_3+k_1k_3)}{k^5}+\frac{120k_1k_2k_3}{k^6})~.
\end{equation}
\begin{itemize}
\item $H_3^{(12)}=\frac{\bar{M}_8H^2\tau^2}{6}\partial_i\partial_i\pi(\partial_j\partial_k\pi)^2$
\end{itemize}
\begin{displaymath}
2\operatorname{Im}\int^0_{-\infty}d\tau\langle0|\pi^3H_3^{(12)}(\tau)|0\rangle=-\frac{2\bar{M}_8}{3}H^2\frac{\cos^2\langle\mathbf{k}_2,\mathbf{k}_3\rangle}{k_1k_2k_3}(\frac{8}{k^3}+\frac{24(k_1k_2+k_2k_3+k_1k_3)}{k^5}+\frac{120k_1k_2k_3}{k^6})
\end{displaymath}
\begin{equation}
+\text{2 perm}.
\end{equation}
\begin{itemize}
\item $H_3^{(13)}=\frac{\bar{M}_9H^2\tau^2}{6}\partial_i\partial_j\pi\partial_j\partial_k\pi\partial_k\partial_i\pi$
\end{itemize}
\begin{displaymath}
2\operatorname{Im}\int^0_{-\infty}d\tau\langle0|\pi^3H_3^{(13)}(\tau)|0\rangle=-2\bar{M}_9H^2\frac{\cos\langle\mathbf{k}_1,\mathbf{k}_2\rangle\cos\langle\mathbf{k}_2,\mathbf{k}_3\rangle\cos\langle\mathbf{k}_1,\mathbf{k}_3\rangle}{k_1k_2k_3}
\end{displaymath}
\begin{equation}
\times(\frac{8}{k^3}+\frac{24(k_1k_2+k_2k_3+k_1k_3)}{k^5}+\frac{120k_1k_2k_3}{k^6})~.
\end{equation}
Notice that the cosine function is used to write the correlators in a more compact form. Using the cosine theorem (note that we have the constraint $\mathbf{k}_1+\mathbf{k}_2+\mathbf{k}_3 = 0$) we can see that this cosine function can be represented by the magnitudes of $\mathbf{k}_1$, $\mathbf{k}_2$, and $\mathbf{k}_3$ without breaking rotational invariance.

In order to know whether the different terms in these expressions for non-Gaussianities (none of which are small) can be canceled by adjusting parameters, we can first add up all the terms and then expand the resulting expression in powers of the momenta. Since we want to obtain solutions which make the result vanish for all values of $\mathbf{k}_i$s, we set all the coefficients to zero, thus obtaining a set of linear equations for the parameters. Because the result is symmetric in $\mathbf{k}_1$, $\mathbf{k}_2$ and $\mathbf{k}_3$, not all of the resulting equations are independent. We finally get ten equations. Solving these equations and considering the coefficients as a vector, there is a solution space with six bases:
\begin{equation}
\mathbf{x}_1=(-1, -\frac{21}{2}, 0, -1, 0, 0, -6, 0, -1, 0, 1)~,
\end{equation}
\begin{equation}
\mathbf{x}_2=(-\frac{2}{3}, -7, 0, -\frac{2}{3}, 0, 0, -4, 0, -1, 1, 0)~,
\end{equation}
\begin{equation}
\mathbf{x}_3=(\frac{1}{6}, \frac{1}{2}, 0, 0, 0, 0, -1, 1, 0, 0, 0)~,
\end{equation}
\begin{equation}
\mathbf{x}_4=(0, -1, 0, 0, 0, 1, 0, 0, 0, 0, 0)~,
\end{equation}
\begin{equation}
\mathbf{x}_5=(0, 0, 0, -1, 1, 0, 0, 0, 0, 0, 0)~,
\end{equation}
\begin{equation}
\mathbf{x}_6=(-\frac{3}{4}, -\frac{9}{8}, 1, 0, 0, 0, 0, 0, 0, 0, 0)~.
\end{equation}
where each $\mathbf{x}_i$ vector corresponds to the effective field theory parameters
\begin{equation}
\mathbf{x}_i=(M_2,M_3,\bar{M}_1,\bar{M}_2,\bar{M}_3,\bar{M}_4,\bar{M}_5,\bar{M}_6,\bar{M}_7,\bar{M}_8,\bar{M}_9)~.
\end{equation}
Combining the "no ghost" conditions and the  obvious requirement on the parameters, for example $M^4_2\geq 0$, we can obtain a set of inequalities for the coefficients of the linear combination of the bases:
\begin{equation}
-6C_1-4C_2+C_3-\frac{9}{2}C_6\geq 0~,
\end{equation}
\begin{equation}
-6C_1-4C_2-C_3\geq 0~,
\end{equation}
\begin{equation}
-6C_1-4C_2-6C_5\geq 0~,
\end{equation}
\begin{equation}
-6C_1-4C_2+\frac{2}{7}C_3-\frac{4}{7}C_4-\frac{9}{14}C_6\geq 0~,
\end{equation}
\begin{equation}
C_3\geq 0~,
\end{equation}
\begin{equation}
C_5\geq 0~.
\end{equation}
where $C_i$ are the coefficients of $\mathbf{x}_i$. Finally, by solving these inequalities, we obtain the range of the parameters where the three-point correlation function vanishes. For example, we set
\begin{displaymath}
M_2=\sqrt[4]{7501}, M_3=\sqrt[4]{\frac{22523}{2}}, \bar{M}_1=\sqrt[3]{\frac{-10000}{H}}, \bar{M}_2=\frac{1}{H}, \bar{M}_3=0, \bar{M}_4=\frac{-1}{\sqrt[3]{H}}~,
\end{displaymath}
\begin{equation}
\bar{M}_5=\frac{\sqrt{6}}{H}, \bar{M}_6=0, \bar{M}_7=\frac{1}{H^3}, \bar{M}_8=0, \bar{M}_9=\frac{-1}{H^3}~.
\end{equation}
For this set of parameters (constants or slowly changing in time) the three point correlation function will vanish for arbitrary $\mathbf{k}_i$s. Therefore, in the regime satisfying the above inequalities, the theory will not produce large non-Gaussianities in the three point function. But outside this regime one will obtain non-Gaussianities of order $\mathcal{O}(1)$ in the three point function.

\end{document}